\documentclass[number, 3p, twocolumn]{elsarticle} 
\usepackage[english]{babel}
\usepackage{amssymb}
\usepackage{mathtools, amsmath, amssymb,bbm}
\usepackage{graphicx}
\usepackage{natbib}

\usepackage{algorithm}
\usepackage{algpseudocode}
\usepackage{siunitx}
\usepackage{caption} 
\makeatletter
\def\ps@pprintTitle{%
 \let\@oddhead\@empty
 \let\@evenhead\@empty
 \def\@oddfoot{}%
 \let\@evenfoot\@oddfoot}
\makeatother

\begin{document}

\title{A Non-Intrusive Load Monitoring Approach for Very Short Term Power Predictions in Commercial Buildings}

\author{Karoline~Brucke}
\author{Stefan~Arens\corref{cor1}}
\author{Jan-Simon~Telle}
\author{Thomas~Steens}
\author{Benedikt~Hanke}
\author{Karsten~von~Maydell}
\author{Carsten~Agert}

\cortext[cor1]{Corresponding author, email adress: Stefan.Arens@dlr.de}
\cortext[cor2]{All authors were with: DLR Institute of Networked Energy Systems, Carl-von-Ossietzky-Str.~15, 26129~Oldenburg, Germany}

\begin{abstract}
This paper presents a new algorithm to extract device profiles fully unsupervised from three phases reactive and active aggregate power measurements. The extracted device profiles are applied for the disaggregation of the aggregate power measurements using particle swarm optimization. Finally, this paper provides a new approach for short term power predictions using the disaggregation data. For this purpose, a state changes forecast for every device is carried out by an artificial neural network and converted into a power prediction afterwards by reconstructing the power regarding the state changes and the device profiles. The forecast horizon is 15 minutes. To demonstrate the developed approaches, three phase reactive and active aggregate power measurements of a multi-tenant commercial building are used. The granularity of data is 1~s. In this work, 52 device profiles are extracted from the aggregate power data. The disaggregation shows a very accurate reconstruction of the measured power with a percentage energy error of approximately 1~\%. The developed indirect power prediction method applied to the measured power data outperforms two persistence forecasts and an artificial neural network, which is designed for 24h-day-ahead power predictions working in the power domain. 
  \end{abstract}

\maketitle

\section{Introduction}
Due to a higher share of renewable energies and the increasing electrification of our society, the electricity grid is facing more challenges such as instabilities and sudden increases in energy supply or demand. 
A possible solution to avoid overloading without a massive increase in grid infrastructure is energy management on both, the supply and the demand side of the electrcity grid \cite{strbac2008}. 
Energy management relies among other things on high quality forecasts of the electricity supply and demand for different time horizons of seconds to months \cite{tran2013,arcos2017,Wan2015}. 
Predictions for seconds to minutes are referred to as \textit{very short term} prediction. 
The size of systems of predictions differ from the high voltage grids to the device level \cite{Pedersen2008,Beccali2004,Li2013}.
Power predictions on the demand side include random human behavior and thus show erratic and highly volatile patterns.
Especially, very short term predictions are highly influenced by randomness and are more difficult to carry out than long term predictions \cite{Lang2019}.   
This behavior is particularly evident in buildings like households and industrial or commercial buildings. 
However, commercial and industrial buildings show a more repetitive demand than households due to the division of working time and non-working time trough for instance shift work and repetitive tasks.
Thus, there is much potential to carry out high quality power predictions in commercial buildings which additionally mostly have a higher electricity demand than households as \cite{UBA2019} shows for Germany.

For power predictions, mostly machine learning methods, in particular artificial neural networks (ANNs), are applied in order to learn interrelations between past and future consumption data \cite{Hippert2001,Raza2015}. 
Power prediction methods in general mainly work directly with the quantity to be predicted and are often not able to predict sudden events resulting in sharp power rises \cite{Lang2019,Raza2015,Alonso2020}.
In this work, we introduce a different approach of power predictions based on non-intrusive load monitoring (NILM), which was firstly described by Hart in 1992 \cite{Hart1992}. 
NILM is also called energy disaggregation because it divides aggregate consumption data into the contributions of single devices. 
Energy disaggregation aims for the description of the state of every appliance in a building without a massive increase in metering infrastructure and was carried out by various methods in the past \cite{Faustine2017,Zoha2012}.
Most disaggregation methods work with model building based on prior knowledge and also labeled data sets as \cite{kolter2011redd} which are mostly not present in reality. 
Therefore, NILM approaches that work without labeled data sets are required. 

Disaggregation produces additional knowledge of the building which can be used for different purposes like power predictions.
In \cite{Welikala2017}, the authors incorporate appliance usage patterns to improve performance of load forecasting and in \cite{Wurm2018} the authors use NILM and a subsequent clustering of similarly behaving appliances as a preprocessing step for their forecast algorithm.
 
In this work, we present a new approach of power predictions using the state data of devices produced by our NILM approach. 
Thus, we develop a \textit{state} prediction of devices and carry out a power prediction by reconstructing the aggregate power from the state data with respect to the according device profiles. 
Therefore, we present an unsupervised NILM approach whose produced state data is used for short term power predictions. 
For this purpose, we firstly state our developed algorithm which is able to extract device profiles from aggregate consumption data unsupervised with methods from machine learning and statistics. 
No prior knowledge of the building is required beforehand to calculate the device profiles. 
Afterwards, we disaggregate the aggregate signal using particle swarm optimization as developed by the authors of this paper and extensively covered in \cite{Brucke2020}. 

\begin{figure}[!t]
\centering
\includegraphics[width=0.45\textwidth]{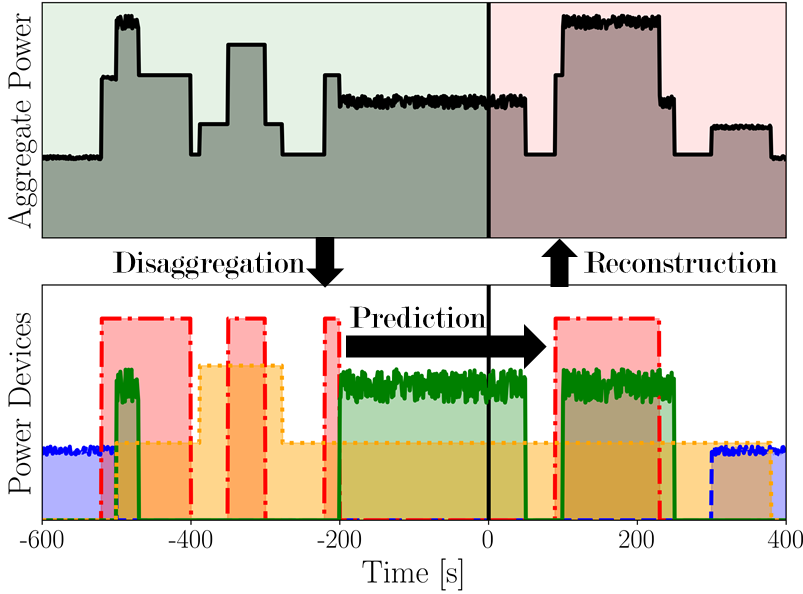}
\caption{Graphical representation of power predictions based on power disaggregation. The aggregate power signal (top) is disaggregated until $t = 0$. This results in the power contributions of different devices (bottom). From $t = 0$ the state of device is predicted and thereafter the aggregate power signal gets reconstructed from state data of single devices}
\label{fig:NILM}
\end{figure}

Figure~\ref{fig:NILM} shows this procedure visually with past aggregate consumption data being disaggregated to the device level. Afterwards the state of devices gets predicted and the aggregate power consumption is reconstructed according to the state prediction. 

In Section~\ref{chap:Data} we present the used data set.
Thereafter, the methodology is outlined in Section~\ref{chap:Methodology}, whereas we start by describing the assumed disaggregation problem in Section~\ref{chap:DisProblem}. 
Afterwards, the developed and used methods are presented including device profile extraction, disaggregation with particle swarm optimization and very short term power prediction using an artificial neural network. 
In order to show results, the developed methods are exemplary applied to the power data of a commercial building in Section~\ref{chap:Results}. After a discussion in Section~\ref{chap:Discussion} of results we finish with a conclusion in Section~\ref{chap:Conclusions}.

\section{Data Description}
\label{chap:Data}
In this work, we use power data of one measuring point in a multi-tenant commercial building as a test data set for our developed methods. 
The granularity is 1\,s. 
The data represents a production facility and workshop and contains six features: Three phases of active and reactive power. 
The six features of the power measurements are referred to as $P_0 \dots P_5$ whereas $P_0 \dots P_2$ represent the three active power phases and $P_3 \dots P_5$ represent the three reactive power phases respectively. 
The data set includes the power measurements from December 1\textsuperscript{st}, 2018 until March 29\textsuperscript{th}, 2019. 
On average there are 0.0023\,\% of data points missing. 
Gaps are filled by the last known value. We use the power analyzer UMG 604 PRO from Janitza Electronics. 
According to the manufacturer the measuring error is less than 0.4\,\% which we neglect in this work \cite{Janitza}. 
Table~\ref{tab:data_analysis} shows four key indicators of the used data set. 

\begin{table}[!t]
\renewcommand{\arraystretch}{1.3}
\caption{Data analysis of the measured power data set}
\label{tab:data_analysis}
\centering
\begin{tabular}{|c|c|}
\hline
Property & Value\\  
\hline
$P_\mathrm{min}$ [kW] & 2.261 \\
$P_\mathrm{max}$  [kW]& 98.95  \\
$\operatorname{Energy}_\mathrm{mean}$ per Day [kWh] & 534.5\\
$P_\mathrm{mean}$  [kW]& 22.27 \\
\hline
\end{tabular}
\end{table}

\section{Methodology}
\label{chap:Methodology}
In the first of the following sections, the assumed formulation of the disaggregation problem is stated. 
Sencondly, the device profile extraction method is presented. 
Thereafter, the used disaggregation method PSO is described briefly in the subsequent section. 
The disaggregation based power prediction procedure is outlined in the last of the following sections. 

\subsection{Formulation of the Disaggregation Problem}
\label{chap:DisProblem}
We assume a very similar formulation of the disaggregation problem as in \cite{Brucke2020}. 
The aggregate power at time $t \in \{ 0, 1, \dotsc, T\}$ called $P(t)\in \mathbb{R}^6$ is assumed to be a linear combination of device profiles according to their state changes as described in the following equation: 

\begin{multline}
    P(t) = \sum_{\substack{i,\tilde t\\s_{i}(\tilde t) = 1}} s_{i}(\tilde t) l_{i}(t + \tilde t) +\\
     \sum_{\substack{i,\tilde t\\s_{i}(\tilde t) = -1}} s_{i}(\tilde t) \mathbbm{1}_{(\tilde t,T)}(t) p_{i} + \epsilon(t)
\label{eq:aggregate}
\end{multline}

\begin{figure}[!t]
\centering
\includegraphics[width=0.45\textwidth, scale = 2]{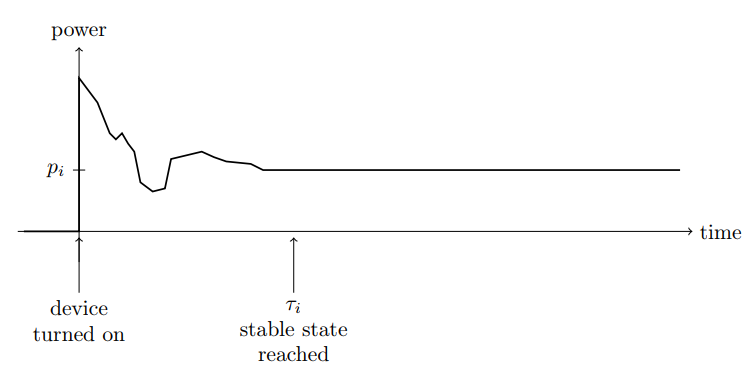}
\caption{Graphical representation of a theoretical device profile including dynamic behavior in the beginning and the stable operating state after $\tau_i$.}
\label{fig:theory_profile}
\end{figure}

The device profile of device $i \in \{0, 1, \dotsc, M\}$ contains a dynamic profile $l_i$ and a power value of the stable operating state $p_{i} \in \mathbb{R}^6$ with $\tau_{i}$ being the (typical) time until this state is reached. This behavior is shown in Figure~\ref{fig:theory_profile}

$S \in \{0,1,-1\}^{T \times M}$ denotes the so-called state-changes-matrix with $s_i(t)$ being the $t$\textsuperscript{th} row and the $i$\textsuperscript{th} column of $S$. 
If $s_i(t) = 1$, the device $i$ is switched on at time $t$ and for $s_i(t) = -1$ the device is switched off respectively.
When $s_i(t) = 0$, the state of device $i$ remains the same. 
$\epsilon(t)$ is referred to as always-on-component or noise. 
Given these assumptions for the aggregate power signal, the following optimization problem has to be solved: 

\begin{equation}
    \min_\mathrm{S} E \bigg( P, P_{\mathrm{S}}  \bigg)
\label{eq:problem}
\end{equation}
$P$ denotes the measured aggregate power signal. 
$P_{\mathrm{S}}$ denotes the reconstructed or approximated power according to Equation~\ref{eq:aggregate} using the state changes matrix $S$ and the device profiles $l_i$.
$E(P,P_{\mathrm{S}})$ represents an error function of $P$ and $P_{\mathrm{S}}$. 
The state changes matrix $S$ and the device profiles $l_i$ have to be found in order to minimize the error $E$.

\subsection{Extraction Procedure of Single Device Profiles}
\begin{figure}[!t]
\centering
\includegraphics[width=0.45\textwidth, scale = 2]{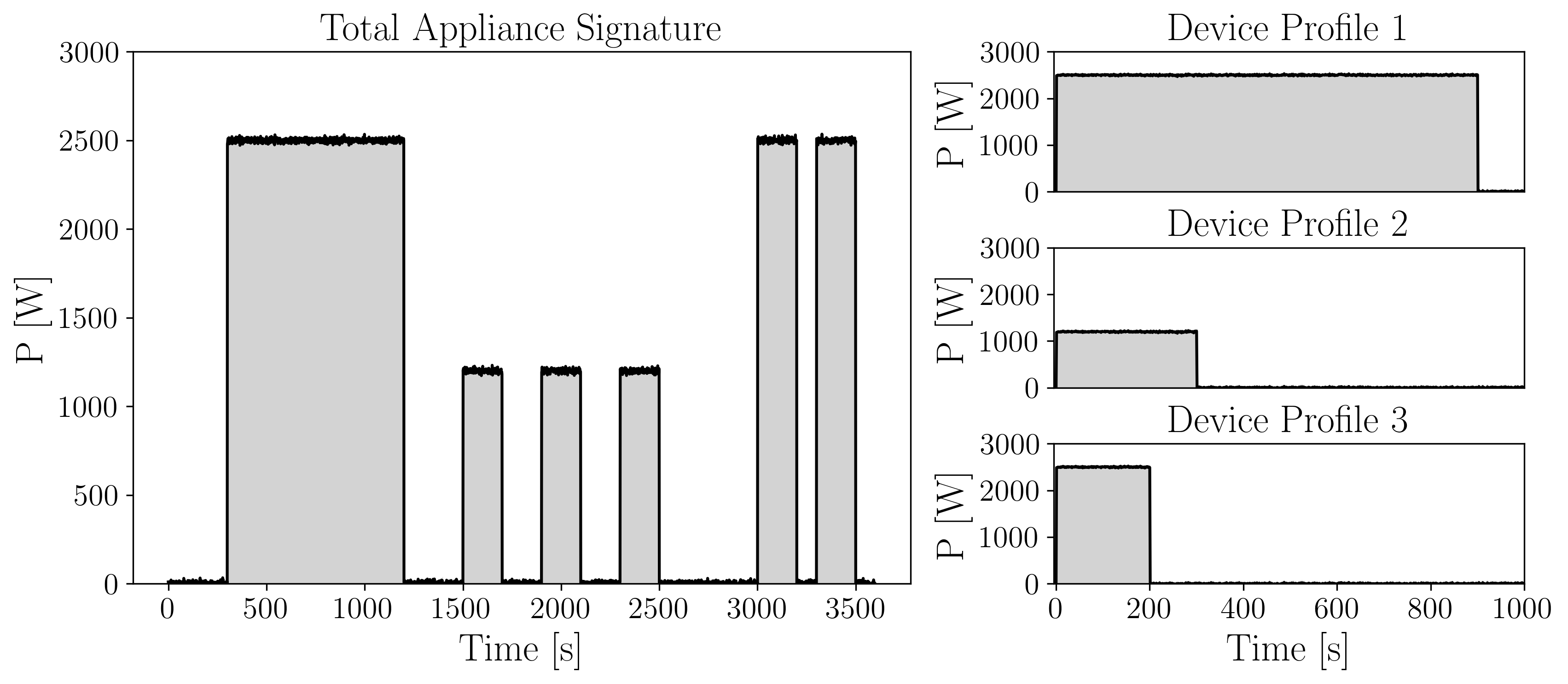}
\caption{Graphical representation of the separation of complex appliance signatures into simple device profiles. The left profile contains repetitive patterns and is divided into three characteristic simpler profiles that represent the characteristic patterns.}
\label{fig:Separation}
\end{figure}

For device profile extraction, we assume a device having a binary state, i.e. the device is only in the state ON or OFF. Stand-by modes or different operational modes of one appliance would be described as individual device profiles. 
This applies for complex programs of some appliances as well. Fig.~\ref{fig:Separation} shows a graphical and generic representation of the division of a complex appliance signature into simplified device profiles. 

The device profile extraction algorithm firstly detects times of events in the aggregate power signal by identifying peaks in the derivative of the aggregate power signal. 
Then, events are clustered using the k-means algorithm to determine the characteristics when switching the specific device types on or off. 
Afterwards, the clusters are cleaned and merged. 
In order to determine the typical run-time of the device, i.e. the length of the device profile, the clusters are split using Gaussian Mixture Models according to the characteristic ON-duration. 
Finally, median blending is used to extract the device profiles from the aggregate power signal.

\subsubsection{Peak Analysis}
We start by identifying when state changes of devices take place. 
For this purpose, we use the derivative of the measured aggregate power signal $P$ which is denoted by $\Delta P : \{1,\dotsc, T-1 \} \to \mathbb{R}^6$  and is calculated according to Equation~\ref{eq:derivative} where $t+1$ denotes the subsequently measured point in time with respect to $t$. 
Due to the measuring frequency of $1~\si{\hertz}$, the relation is simplified to: 
\begin{equation}
 \Delta P(t) = \frac{P(t+1)-P(t)}{(t+1)-t} = \frac{P(t+1)-P(t)}{1\,\mathrm{s}}
 \label{eq:derivative}  
\end{equation}
We assume that a state change takes places, when a sharp increase or decrease in the measured power is observable. 
These inflection points in the aggregate power signal result in maxima or minima in the derivative. Maxima are referred to as ON-events and minima as OFF-events in the following. 
In order to identify events, we take the sum of active phases $P_\mathrm{tot} \in \mathbb{R}^{T}$ with $P_\mathrm{tot} = P_1 + P_2 + P_3$ into account. We perform a peak analysis in the derivative of the sum of the three phases of active power $\Delta P_\mathrm{tot}$ with $\Delta P_\mathrm{tot} \in \mathbb{R}^{T-1}$. 
For the peak analysis, we take all values of $\Delta P_\mathrm{tot}$ into account, which are above a threshold value $\varepsilon_\mathrm{threshold}$, thus $|\Delta P_\mathrm{tot}(t)| \geq \varepsilon_\mathrm{threshold}$. 
The threshold can to be chosen with respect to the given power data. 
The choosing process of a peak threshold could be automated in the future. 
We assume, that the process of switching on or off a device is completed within $1~\si{\sec}$. 
When $\Delta P_\mathrm{tot}(t)$ is an event, we denote the respective time by $t_\mathrm{p}$ and call $t_\mathrm{p}$ event-time. 
We introduce the following peak criterion which defines time $t$ to be or not to be an ON-event time $t_\mathrm{p}$:
\begin{multline}
t = t_\mathrm{p} \Leftrightarrow 
\Delta P_\mathrm{tot}(t-1)<\Delta P_\mathrm{tot}(t)  \wedge  \\
\Delta P_\mathrm{tot}(t+1)<\Delta P_\mathrm{tot}(t) \\
\wedge \Delta P_\mathrm{tot}(t) \geq \varepsilon_\mathrm{threshold}
\label{eq:peak_criterion}  
\end{multline} 
Equation \ref{eq:peak_criterion} accordingly applies for OFF-events with reversed signs. 
The set of $N$ events is referred to as $D = \{\Delta P(t_\mathrm{p,1}), \dots , \Delta P(t_{\mathrm{p},N})\}$.

\subsubsection{Cluster Analysis of Events}
The relation between active and reactive power shows to be distinctive for a specific device type \cite{Ng2009}. 
Therefore, we assume the increase or decrease in the three phases of active and reactive power at the time of an event to be characteristic for the specific device type.
With this assumption, we can cluster the extracted events $\Delta P_\mathrm{tot}(t_\mathrm{p})$ according to their values in all six power features to distinguish the device types. 
For the cluster analysis we use the well known k-means cluster algorithm. 
It is assumed that the specific patterns of an ON-event correspond to those of an OFF-event with reversed signs. 
Clustering is therefore only performed for the ON-events and the OFF-events are assigned to the cluster centers with reversed signs with the smallest deviation afterwards.
The k-means cluster algorithm divides a given data set $D = \{\Delta P(t_\mathrm{p,1}), \dots , \Delta P(t_{\mathrm{p},N})\}$ into $K$ Clusters in such a way, that the Euclidean distance of each data point to the nearest cluster center is minimized. The number of clusters $K$ has to be given. 
This can be formalized as: 

\begin{equation}
\min_{r_{nk}, \vec{c}_k} \quad \sum_{n=1}^{N} \sum_{k=1}^{K} r_{nk} |\Delta P(t_{\mathrm{p},n}) - \vec{c}_k|^2
\label{eq:Kmeans}
\end{equation} 

$r_{nk} = 1$ if the event $\Delta P(t_{\mathrm{p},n})$ belongs to the cluster $k$ and $r_{nk} = 0$ for all other clusters. 
Cluster centers are denoted by $\vec{c}_k \in \mathbb{R}^{6}$ and the according cluster is a set of assigned events denoted by $c_k$. 
The k-means cluster algorithm solves the minimization problem using the expectation-maximization-method \cite{Bishop2006a}.
In order to determine the optimal number of clusters $K_\mathrm{opt}$ for the given events, the Calinski-Harabasz-Score ($\operatorname{CH}$) is used, which is defined by \cite{Calinski1974}:
\begin{equation}
\operatorname{CH} = \frac{N-K}{K-1} \frac{\sum_{c_k \in C} |c_k||\vec{c}_k -\vec{D}|^2}{\sum_{\vec{c_k}} \sum_{\Delta P(t_{\mathrm{p},i}) \in c_k}|\Delta P(t_{\mathrm{p},i}) -\vec{c_k}|^2}
\label{eq:CalinskiHarabasz}
\end{equation} 
$N$ denotes the number of events. 
The center of the whole data set $D$ is denoted by $\vec{D}$ and $C$ denotes the set of clusters $c_k$. 
The cardinality of cluster $k$ is denoted by $|c_k|$. 
The $\operatorname{CH}$ gets maximum for the optimal $K$ and calculates a ratio between the separation of the clusters and the compactness within each cluster.  
It is multiplied by the pre-factor $\frac{N-K}{K-1}$ to prevent overfitting, because a larger number of clusters $K$ must not always result in a higher value of the $\operatorname{CH}$ than a smaller number of cluster.  

In order to obtain $K_\mathrm{opt}$, we perform a k-means clustering for $K \in \{1\dots 50\}$ and calculate $\operatorname{CH}$ every time. 
We choose 50 as the upper limit to confine the computing time. 
An adaptive method to increase $K$ until the $\operatorname{CH}$ is decreasing again would be possible as well. 

After the first clustering of the extracted events, we perform a cleaning step of the clusters analogously to \cite{Ng2009}.
For this, we define outlier events $\Delta \tilde{P}(t_\mathrm{p})$ to be out of a $2\sigma$-area of the respective cluster where $\sigma$ denotes the standard deviation of the respective cluster.  
All outliers get clustered again with fixed $\tilde{K} = 10$. 
A second $\operatorname{CH}$-analysis would be possible for the outlier events as well but this step is simplified since this cleaning step is optional in the procedure of the extraction of device profiles. 
With the presented clustering procedure, the characteristic increase or decrease in all six power features when switching a device on or off is known.

\subsubsection{Merging Clusters}
In order to improve the clustering of the extracted events, we perform a merging step of clusters based on a similarity measure. 
The similarity of two clusters is evaluated by means of the Pearson correlation coefficient $\rho \in [-1,1]$ and absolute percentage error (APE) calculated for every combination of two cluster centers. 
The Pearson correlation coefficient of two cluster centers $\vec{c}_i$ and $\vec{c}_j$ is defined by the following equation \cite{McClarren2018}: 

\begin{equation} 
\rho (\vec{c}_i,\vec{c}_j) = \frac{\sigma_{\vec{c}_i,\vec{c}_j}}{\sigma_{\vec{c}_i} \sigma_{\vec{c}_j}}
\label{eq:Pearson}
\end{equation}
where $\sigma_{\vec{c}_i,\vec{c}_j}$ denotes the co-variance of $\vec{c}_i$ and $\vec{c}_j$. 
The APE is defined by the following equation: 
\begin{equation}
  \operatorname{APE}(\vec{c}_i,\vec{c}_j) = \frac{|\vec{c}_i-\vec{c}_j|}{|\vec{c}_i|}  
 \label{eq:APE}  
\end{equation}
If $\rho(\vec{c}_i,\vec{c}_j)$ and $\operatorname{APE}(\vec{c}_i,\vec{c}_j)$ are above/below a given threshold, cluster $i$ and $j$ are merged. 
For this, a new cluster is created and the cluster members of cluster $i$ and $j$ are assigned to this new cluster with the cluster center $\vec{c}_{i,\mathrm{new}} = 1/2 \cdot (\vec{c}_i + \vec{c}_j)$. 
This calculation of the new cluster center is also applied if the cardinalities of $c_i$ and $c_j$ are different.   

The thresholds are chosen to be: 
\begin{equation}
\label{eq:KorrBedingung}
\rho(\vec{c}_i,\vec{c}_j) > 0.9 \quad  \wedge  \quad APE(\vec{c}_i, \vec{c}_j)< 0.1 
\end{equation}
It is possible, that cluster $c_i$ satisfies the condition in Eq.~\ref{eq:KorrBedingung} with multiple other clusters. 
If that situation applies, $c_i$ is only merged with the cluster of highest similarity.
The newly created cluster $c_{i,\mathrm{new}}$ is not merged again with other clusters. 

\subsubsection{Determination of Run-Time with Gaussian Mixture Models}
For the calculation of device profiles, the typical run-time i.e. time in the state ON, is required. 
Therefore, we determine the time between an ON-event in a specific cluster and the next OFF-event in that cluster for every ON-event in that cluster. 
We perform this calculation for every cluster of events. 
The calculated time is referred to as ON-duration in the following. 
If there are more ON-events than OFF-events, we neglect the surplus ON-events and vice-versa.
For every cluster we present all determined ON-durations in a frequency distribution and observe multiple maxima at different times. 

\begin{figure}[!t]
\centering
\includegraphics[width=0.45\textwidth]{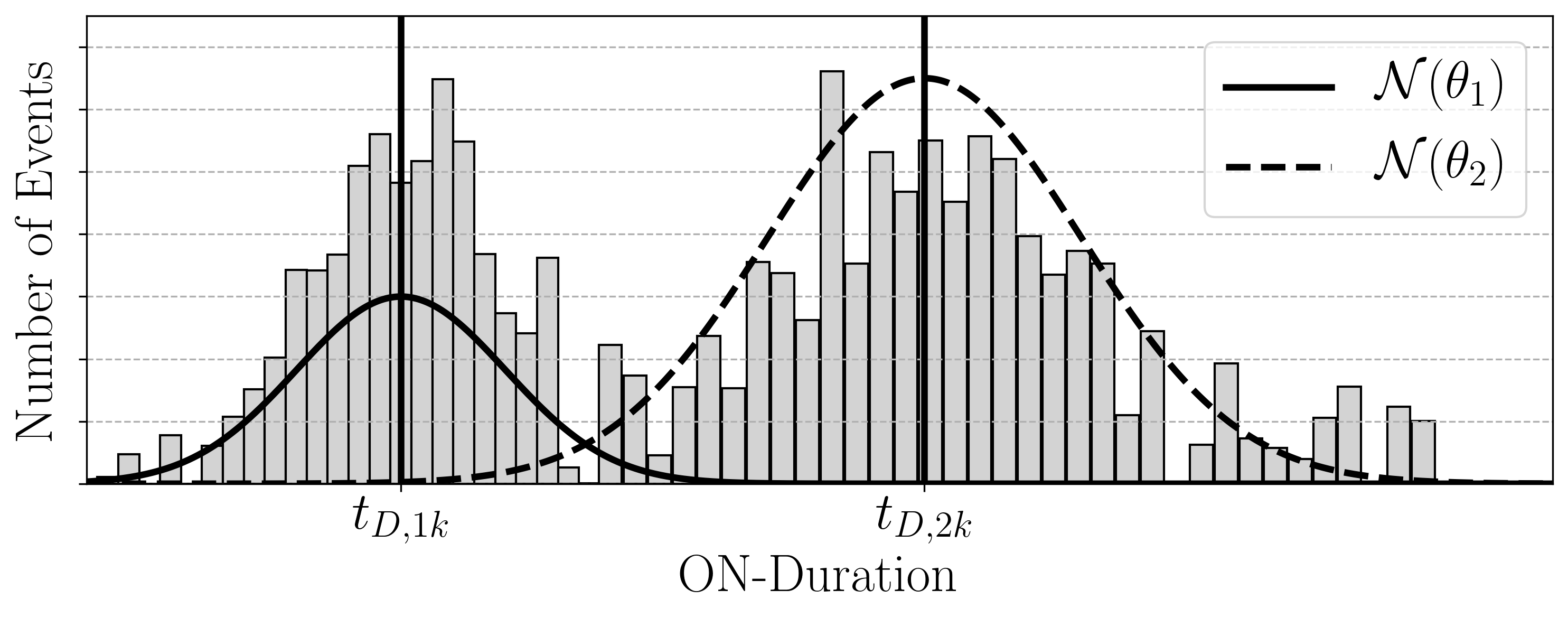}
\caption{Graphical representation of the division of an ON-duration distribution with Gaussian Mixture Models.}
\label{fig:GMM_theory}
\end{figure}

In reality, the ON-duration depends on the kind of use of the individual device, for example if a device is capable of running different programs or if the same device type is used for different tasks. 
In this work, Gaussian Mixture Models (GMMs) are used to divide clusters according to characteristic ON-durations within a specific cluster. 
Figure~\ref{fig:GMM_theory} shows an exemplary distribution of the ON-duration distribution of a cluster with a fitted GMM which divides the distribution in two sub-distributions.  

GMMs determine the properties of sub-distributions in an overall distribution, using only observations of the overall distribution $B = (\vec{x}_1, \dots , \vec{x}_N)$ \cite{Bishop2006c}. 
The a posteriori probability for the GMM is calculated as follows: 
 
\begin{equation}
p(\theta |B) = \sum^{m}_{i=1} \pi_{i} \mathcal{N}(\vec{x}|\vec{\mu}_{i}, \Sigma _{i})
\label{eq:GMM}
\end{equation}
where $p(\theta |B)$ describes the probability of the model parameters $\theta$ given the data set $B$.
The parameters $\theta_i = (\pi_{i} , \vec{\mu}_{i}, \Sigma_{i})$ denote the mixing coefficients, the mean values and covariance matrices of the $i$\textsuperscript{th} of $m$ Gaussian distributions. 
The mean values of the Gaussian distributions represent the mean ON-duration, which will be referred to as $d$ in the following.
Therefore, the ON-duration of device $i$ is denoted by $d_i$.  
The number of sub-distributions $m$ has to be given beforehand. 
The \textit{maximum-likelihood}-method is used together with the \textit{expectaion-maximization}-algorithm to obtain an optimal estimation of $\theta$ \cite{Bishop2006b}.
In order to determine the optimal number of Gaussian distributions $m_\mathrm{opt}$ in the GMM of each cluster, the Bayesian Information Criterion ($\operatorname{BIC}$) is used. The BIC is a measure for comparing different models.
It is defined by the following equation \cite{Bishop2006}:

\begin{equation}
\operatorname{BIC} \approx \frac{1}{2}M \ln N - \ln p(B|\theta) 
\label{eq:BIC}
\end{equation}
$N$ denotes the number of data points in data set $B$ and $M$ the number of parameters in $\theta$. 
According to this definition, the $\operatorname{BIC}$ is to be minimized. 
As soon as $\Delta \operatorname{BIC} > 2$ for two subsequent models $\mathcal{M}_m$ and $\mathcal{M}_{m+1}$, the model $\mathcal{M}_{m+1}$ is selected and the corresponding $m$ is called $m_\mathrm{opt}$. 
The limit for $\Delta \operatorname{BIC}$ to select $m_\mathrm{opt}$ has to be determined empirically. 
In general, $m$ should be increased as long as $\Delta \operatorname{BIC}$ is negative for two subsequent models. 

Given this procedure, every cluster $k$ is divided in $m$ \textit{groups}. The groups that emerge from one cluster share the cluster center (the characteristics at an ON-event and OFF-event) but differ in their characteristic ON-duration. 
An event is assigned to a group if the associated Gaussian distribution is maximum for the ON-duration of this event. 
From in total $K$ clusters emerge $M = \sum^{K}_{k=1} m_{\mathrm{opt},k}$ groups which will be denoted as $G_i$. The ON-duration of $G_i$ is referred to as $d_i$.

\subsubsection{Median Blending}
For the final calculation of the device profiles, median blending is used for all groups. 
Median blending is a method of noise reduction which we will use in order to reduce the \textit{noise} of the aggregate power signal to isolate the device profile from this background \cite{Amri2010}.
For every element $\Delta P(t_\mathrm{p}) \in G_i$ we store and normalize the aggregate power signal from $t_\mathrm{p} \dots t_\mathrm{p} + d_i$. 
The normalized power signal is denoted by $P_\mathrm{norm}$. 
Normalization is carried out by dividing by the maximum power value in the stored part of the aggregate power signal. 
Then, the median for every point in time in the saved aggregate power signal is calculated in all six power features. 
In order to scale the normalized profile back to absolute power values, we use the cluster center of the respective cluster. 
It represents the characteristic increase in power per second when switching on a specific device type. 
Therefore, we integrate the cluster center by multiplying with one second. 
Finally, we scale back the median values by multiplying $\vec{c}_k$ and the normalized $l_i$. 
We define the power profile of device $i$
\begin{equation}
	l_i : \{ 1, \dotsc, d_i \} \to \mathbb{R}^6 \,,
\end{equation}
where $d_i$ denotes the ON-duration, by
\begin{multline}
	l_i(t) = \vec{c}_k \cdot \operatorname{median} \{ P_\mathrm{norm} (t_\mathrm{p} + t) \vert \\
	\text{$t_\mathrm{p}$ is an ON-event of the device} \}
\label{eq:median_blending}
\end{multline}
for every $t \in \{ 1,\dotsc,d_i \}$.
Prerequisite for this procedure are enough events in $G_i$ in order to reduce the noise of the aggregate power signal.

\begin{figure}[!t]
\centering
\includegraphics[width=0.45\textwidth]{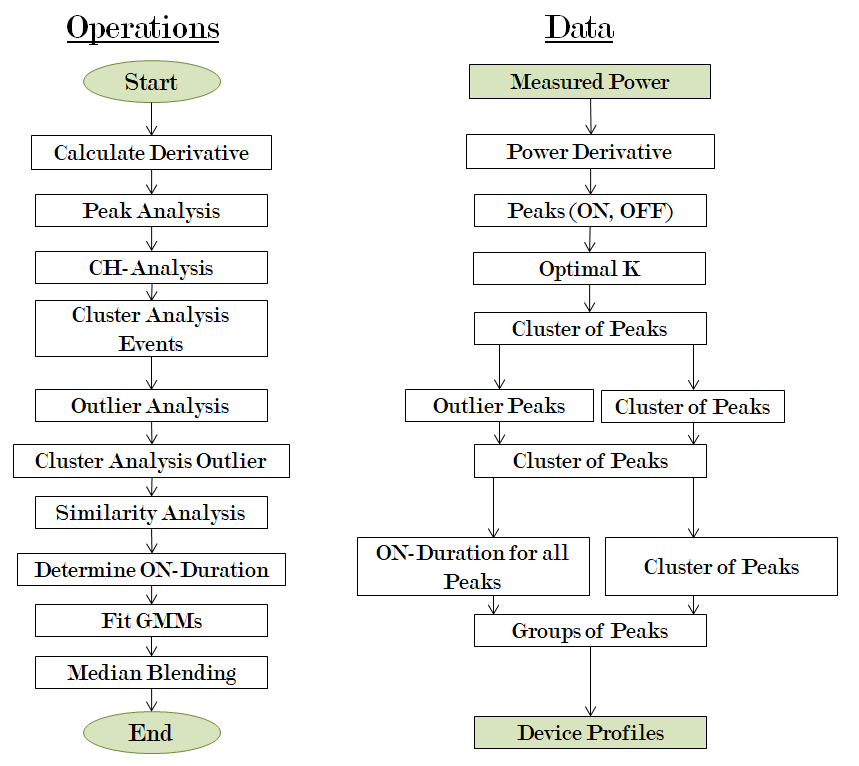}
\caption{Graphical representation of the developed algorithm for device profile extraction. On the left, the operations are depicted while the data of every step in the algorithm is presented on the right.}
\label{fig:time_features}
\end{figure}

\subsection{Disaggregation Procedure}
The disaggregation is carried out by particle swarm optimization (PSO) as described in \cite{Brucke2020} which is a for disaggregation improved version of the original description of PSO by Hart in \cite{Hart1992}.  
The PSO is a metaheuristic used for multidimensional optimization problems as the above presented disaggregation problem.
In this work, we use PSO to determine the state changes matrix $S$.
For this purpose, the extracted device profiles are used.  
The PSO aims for minimizing the following error measure \cite{Brucke2020}: 
\begin{multline}
E^{[a,b)}(P,P_\mathrm{S}) = \alpha \cdot \sum_{t=a}^{b-1} (\vec{P}_{\mathrm{S}}(t)-\vec{P}(t))^2 +\\
 \beta \cdot \sum_{t=a}^{b-2} (\Delta \vec{P}_{\mathrm{S}}(t)-\Delta \vec{P}(t))^2
\label{eq:error_disaggregation}
\end{multline}
with $\alpha +\beta  = 1$ weighting the two summands.  
The algorithm we use in this work to carry out the disaggregation is extensively presented in \cite{Brucke2020}. 
In \cite{Brucke2020} is assumed that a device profiles consists of transient or dynamic behavior and a stable state reach after a specific time $\tau$. 
Thus, we assume the extracted load profiles to represent the dynamic behavior of the device. 
The power value of the stable state is assumed to be the last non-zero power value of the specific device profile.

\subsection{Very Short Term Power Prediction}
In this work, we present a new method for power forecasts that is based on a forecast of state changes of unknown devices. 
The power forecast is carried out by reconstructing the state changes forecast according to Equation \ref{eq:aggregate}. 
The weekends of the used data set show very regular power curves with highly repetitive patterns. 
Thus, it is assumed, that persistence forecasts are sufficient for weekend days.   
Therefore, we only consider working days since they show more complex power demands with many sharp increases and decreases which we are aiming for to predict.
For this purpose, we use an artificial neural network (ANN).  
ANNs have been widely used as a very powerful method for time series prediction in different fields regarding the power grid \cite{al2019,kim2019,torabi2019}. 
Especially, for load and energy forecasts ANNs are preferred due to the non-linearity and randomness within power data \cite{Lang2019}. 
We are aiming for the ANN learning an interrelationship between the last hour of state changes and the state changes within the next 15 minutes.  
We use a feed forward, fully connected ANN based on the supported models and functions of \texttt{keras} \cite{chollet2015}.
In the following, the feature selection and the hyperparameter optimization of the ANN are described. Finally, the training procedure of the ANN is outlined.

\subsubsection{Feature Selection}
The feature selection for the ANN determines the input and target data, also called output data. 
All inputs and outputs have to be normalized to a range of $-1  \dots 1$ so that no feature is weighted more than another during training. 
We use past data of state changes of every device type as inputs and future state changes data as target data. 

Additionally, we introduce three time features:
The first two time features are the sine and cosine function as presented in Figure~\ref{fig:time_features} in the left and the middle illustration. 
The third time features represents the day of week:
We assign a value to every day from Monday (0) to Friday (1) as shown in the right illustration in Figure~\ref{fig:time_features}.

\begin{figure}[!t]
\centering
\includegraphics[width=0.45\textwidth]{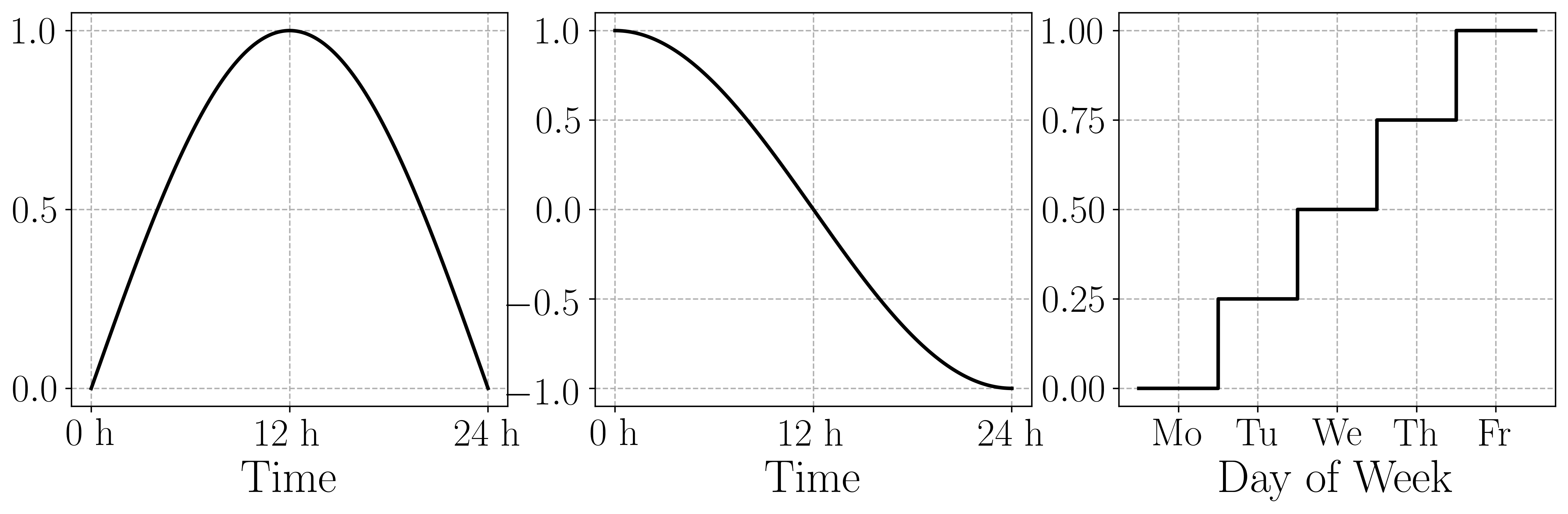}
\caption{Graphical representation of the time features given the ANN as input.}
\label{fig:time_features}
\end{figure}

In order to select input data describing the state changes of every device, we take the data of the previous disaggregation procedure. 
Thus, we select state changes data from $t = -3600\,s \dots t = 0\,s$ to be input data for a state changes prediction for $t = 0\,s \dots t = 900\,s$. 
Secondly, we chose the data from $t = 0\,s \dots t = 900\,s$ of seven days before as input data. 
Thus, for a prediction on a Thursday at 11~am, the state changes data from Thursday 10~am - 11~am as well as the state changes data from the Thursday of one week before from 11~am - 11:15~am are selected.
This feature selection has proven to be helpful, due to the regularity in industrial and commercial data related to the weekday \cite{steens2020}. 
For $M$ given device types, the input data set contains $2M+3$ columns. 
The target data contains only the future state changes data of the $M$ devices. 
Thus, there are $M$ columns in the target data set. 
The number of rows is determined by the size of the training data set, thus the number of rows corresponds to the number of time steps in the training data set. 

\begin{figure}[!t]
\centering
\includegraphics[width=0.45\textwidth]{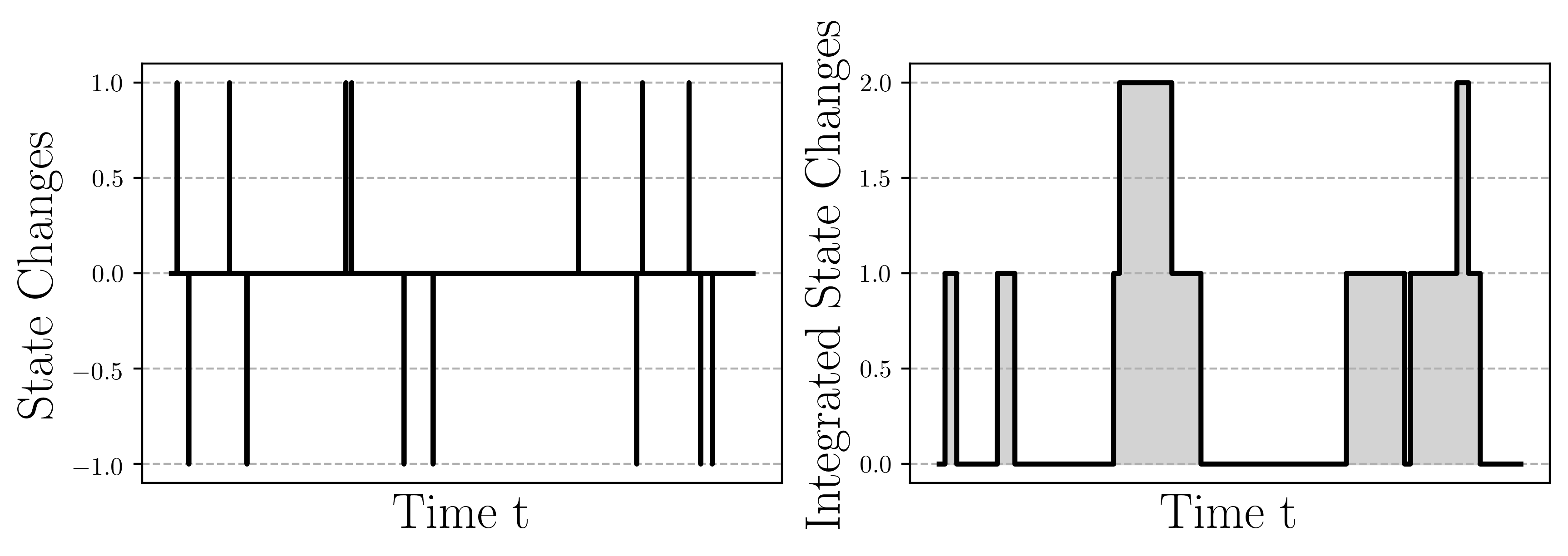}
\caption{Graphical representation of the integration of state changes for the data preparation for the training of the ANN.}
\label{fig:integrate}
\end{figure}

During training, the difference between output data and target data is quantified by calculating an error measure. 
A large proportion of the state changes data is zeros. 
Thus, there is a local optimum in the error measure of the ANN to only predict zeros so no state changes at all.
Therefore, we perform an additional preprocessing step for the state changes data: 
We transform the state changes data to state data via integration. 
Figure~\ref{fig:integrate} shows a graphical representation of the integration procedure of state changes. 
Essentially, the state changes are added up for every device.  
This step bypasses a data structure that contains many zeros.
After integration, the state data is normalized to the range of $[-1 \dots 1]$. 
The integration step applies for the input data as well as for the target data.  
This transformation showed an improved learning process of the ANN during training in comparison to the learning of state changes data. 

\subsubsection{Hyperparameter Optimization} 
The hyperparameter optimization was carried out with the help of \texttt{talos} and the supported random search \cite{talos2019}.
Hyperparameters are all parameters of an ANN that are not adapted during training but have to be set beforehand. 
The following hyperparameters are considered for optimization: 
The \textit{number of neurons} in the hidden layers sets the width of the ANN whereas the \textit{number of hidden layer} determines the depth of the ANN. The number of neurons in the hidden layers has not to be the same for all layers, thus the width of the ANN can vary. 
\textit{Dropout} describes the percentage of neurons that are neglected randomly in every hidden layer during a training step in order to increase the robustness and decrease over-fitting of the ANN \cite{Srivastava2014}.
The \textit{learning rate} is a measure for the step size made in the training process in every iteration. 
A larger learning rate decreases the training time but increases the risk of not fully converging into an error minimum and vice-versa for smaller learning rates. 
The \textit{batch size} determines the number of samples of the training data set that are processed at once. 
Thus, the parameters of the ANN are not adapted after every single sample passed the ANN but only after as many samples as the batch size.
As activation function the \textit{relu} function proved to have the best outcome in this work.  
The chosen values of the optimized hyperparameters are presented in Table~\ref{tab:ANN_Params}.  

\begin{table}[!t]
\centering
\caption{Hyperparameter and their chosen values for the state forecast using an ANN}
\begin{tabular}{|l|l|}
\hline
Hyperparameter Name & Selected Value \\ 
\hline
Neurons in hidden layer & 214 \\
Number of hidden layers & 3 \\
Dropout & 5\,\% \\
Learning rate & 0.01 \\
Batch size & 2048 \\
Activation function & Relu\\
\hline
\end{tabular}
\label{tab:ANN_Params}
\end{table}

Given these hyperparameters, the chosen model has 137868 trainable parameters.   
 
We chose the mean squared logarithmic error ($\operatorname{MSLE}$) as error measure which is defined as follows: 
\begin{multline}
\operatorname{MSLE}(y_\mathrm{target}, y_\mathrm{Out}) = \\
\frac{1}{N} \sum^N_{i=0} \Biggl(\frac{\log(y_{\mathrm{target},i}+1)}{\log(y_{\mathrm{Out},i}+1)}\Biggr)^2
\end{multline}  

Due to the logarithmic character of the $\operatorname{MSLE}$, it penalizes deviations at small values more heavily than error measures like root mean squared error or mean absolute error.  
This showed an improved training process given the structure of data in this work. 
In order to evaluate and compare the results of prediction, we use the same error measures as for the validation of the disaggregation results.

\subsubsection{Training of the Neural Network}
The training is performed with an Intel i7-6700k processor, 16GB of RAM and a Geforce GTX 1050 graphic card with 768 CUDA cores. 
The data set for training includes 55 working days from January 2019 to March 2019, thus it contains 4752000 rows and $2M+3$ columns. 
During training, $95\%$ of the data get used for training the network and $5\%$ get used as a validation data set. 
As soon as the error on the independent validation set increases, the training is stopped. 
This is carried out by the early stopping option of \texttt{keras} \cite{chollet2015}. 
As a postprocessing step, we calculate the derivative, thus the reverse procedure of the shown itegration.
The output values of the used ANN are floats and not integers as assumed in Equation \ref{eq:aggregate}. 
Thus, we interpret the outputs as probabilities of state changes of the devices. 
In order to reconstruct the power, we allow floats and calculate a weighted sum and not a discrete sum. 
Therefore, Equation \ref{eq:aggregate} changes as follows: 

\begin{multline}
    P(t) = \sum_{\substack{i,\tilde t\\s_{i}(\tilde t) > 0.1}} s_{i}(\tilde t) l_{i}(t + \tilde t) +\\
     \sum_{\substack{i,\tilde t\\s_{i}(\tilde t) <-0.1}} s_{i}(\tilde t) \mathbbm{1}_{(\tilde t,T)}(t) p_{i} + \epsilon(t)
\label{eq:aggregate_float}
\end{multline}
with $s_{i}(t) \in \mathbb{R}$. 
We define a threshold of 0.1 to take an element of the prediction into account for reconstruction. As always-on-component $\epsilon$, we give each short term prediction the last measured power value. Thus, for a prediction from $t = 0$ to $t = 900$ we set $\epsilon = P(-1)$.

\section{Results}
\label{chap:Results}
In this section we present the results of the application of the developed methods to the above described data set. 
For the device profile extraction we use the data from January and February 2019. 
Thereafter, we disaggregate the whole data set. 
In order to train the forecast algorithm, we use data from January until March 2019. 
The testing of the forecast algorithm is carried out using the last two days in the data set: 28\textsuperscript{th} and 29\textsuperscript{th} March, 2019. 
Since the forecast horizon is 15~min, we are able to perform an evaluate 188 single power predictions on the test data set. 
In order to validate the results of disaggregation and prediction, we use the error measures root mean squared error (RMSE), mean absolute error (MAE), mean absolute percentage error (MAPE) and the percentage energy difference ($\operatorname{Energy}_\mathrm{E}$) as in \cite{Brucke2020}.

\subsection{Device Profile Extraction}
\begin{figure}[!t]
\centering
\includegraphics[width=0.45\textwidth]{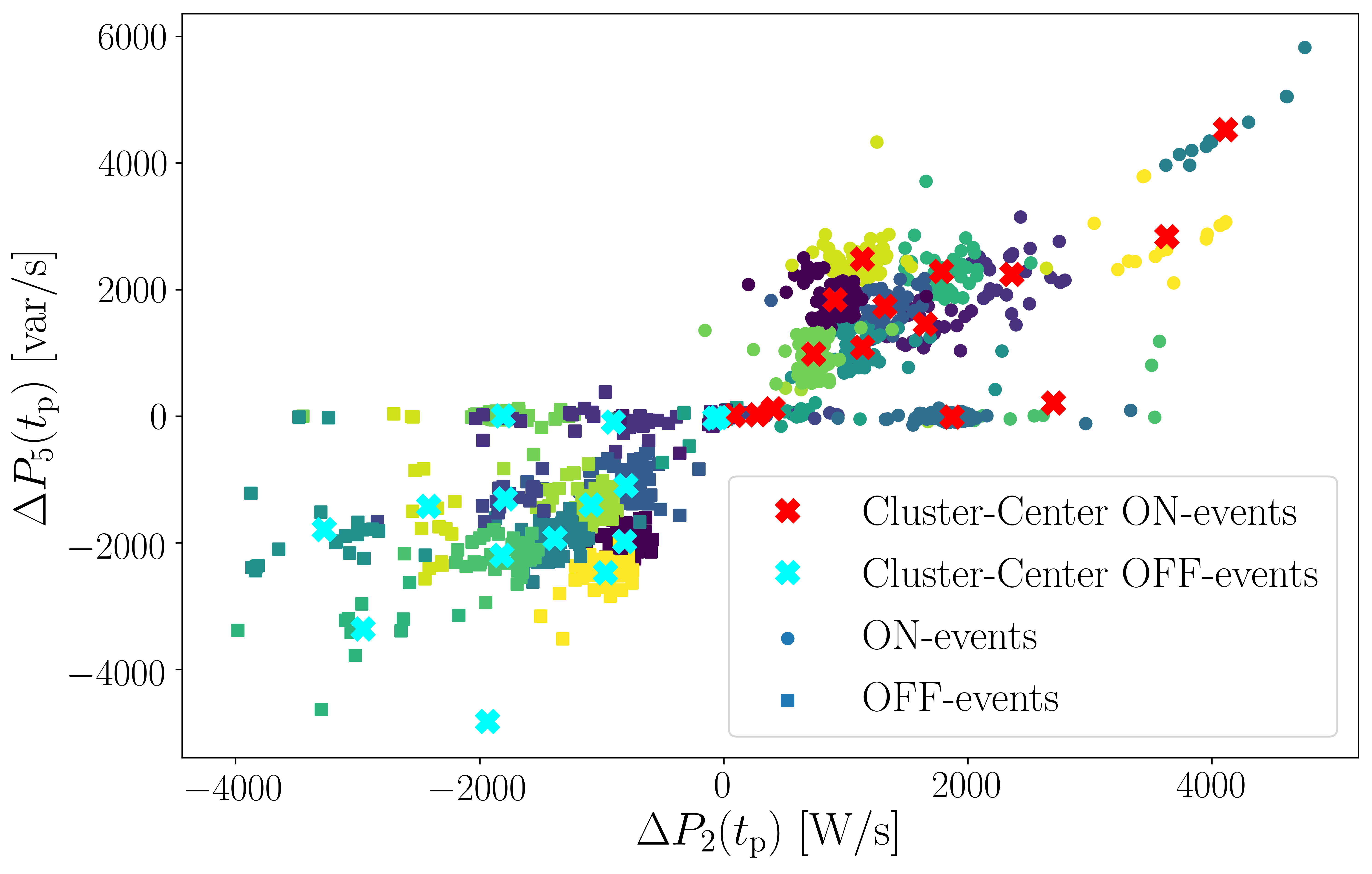}
\caption{Clustering of ON-events and OFF-events individually for December 4\textsuperscript{th}, 2018. Presented are two of six power derivative features.}
\label{fig:Cluster_Analysis}
\end{figure}

\begin{figure}[!t]
\centering
\includegraphics[width=0.45\textwidth]{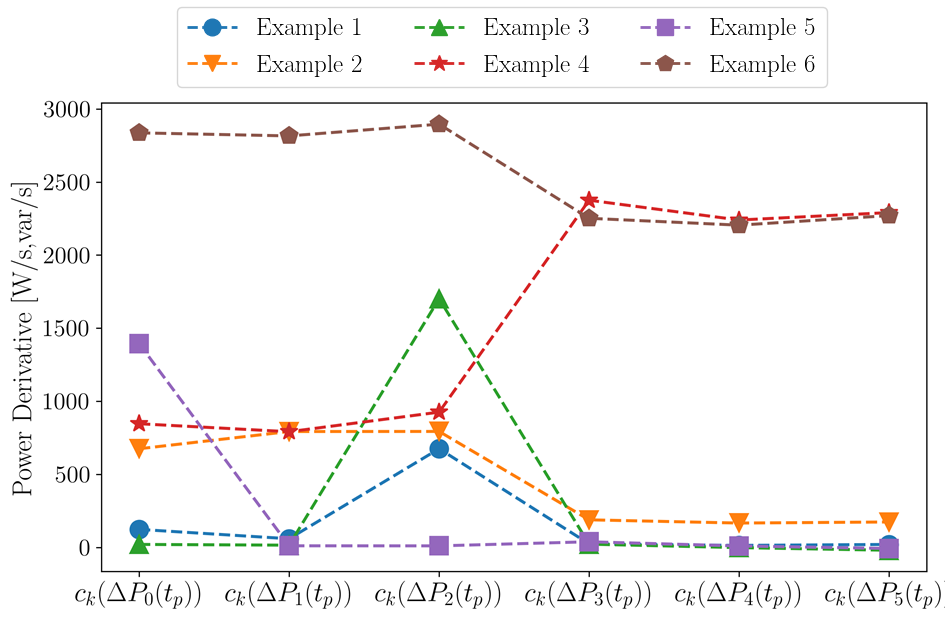}
\caption{Six examples of cluster centers and their characteristics in all features of the power derivative.}
\label{fig:Centers}
\end{figure}

In Figure~\ref{fig:Cluster_Analysis} an exemplary cluster analysis of one day of data (December 4\textsuperscript{th}, 2018) is  shown for the elements $\Delta P_2(t_\mathrm{p})$ and $\Delta P_5(t_\mathrm{p})$. 
ON-events are depicted as well as OFF-events and the symmetry to the central point zero is clearly visible.  
It is apparent, that the relation of the increase in active and reactive power is not randomly distributed, but forms clusters. 
ON-events and OFF-events are clustered individually.
The cluster forming behavior becomes clearer taking all six features of the power derivative into account. 
Therefore, all six features of six exemplary cluster centers are presented in Figure~\ref{fig:Centers}.

It is visible, that the clusters have very distinct characteristics regarding the relation between the six features. 
Whereas Example 1, 3 and 5 only show an increase in one phase of power, the other four examples seem to represent three-phase connected devices. 
They approximately have the same power derivative at an ON-event in all three phases in active and reactive power. 
The relation between active and reactive power is very distinct. 
While the Examples 1, 3 and 5 show almost no increase in reactive power when switched on, Example 4 and 6 have significant reactive power increases. 
For Example 4 the increase in reactive power is even higher than the increase in active power. 
 
\begin{figure}[!t]
\centering
\includegraphics[width=0.45\textwidth]{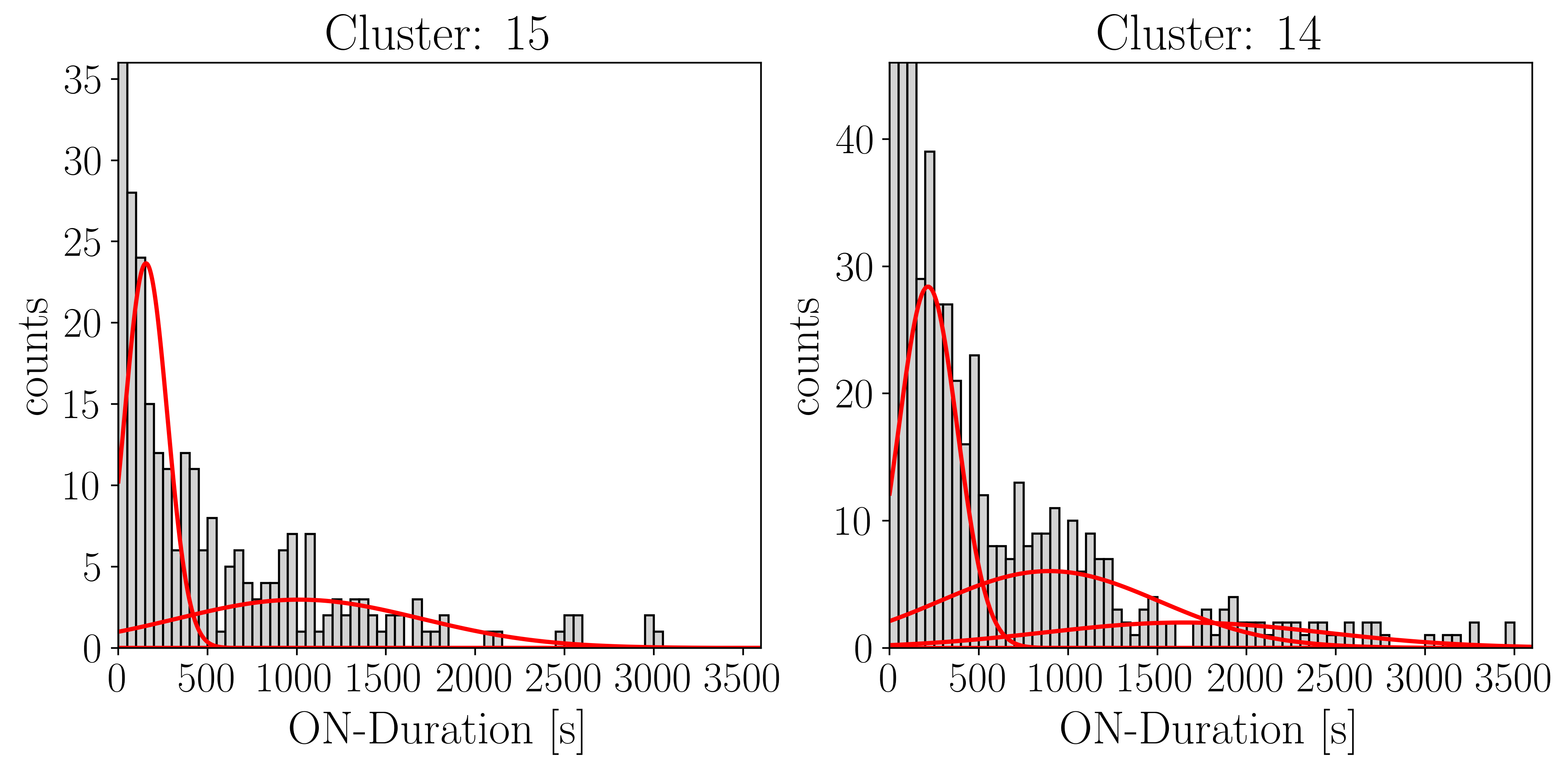}
\caption{Two examples of GMMs for ON-duration distributions}
\label{fig:GMM_example}
\end{figure}

To show the separation of clusters according to their ON-duration, Figure~\ref{fig:GMM_example} shows two exemplary ON-duration distributions with the respective fitted GMMs. 
Cluster 15 from Figure~\ref{fig:GMM_example} gets divided into two groups with approximate ON-durations of 200\,s and 1000\,s. 
On the other hand, Cluster 14 gets divided into three groups with approximate ON-durations of 250\,s, 900\,s and 1900\,s.

\begin{figure}[!t]
\centering
\includegraphics[width=0.45\textwidth]{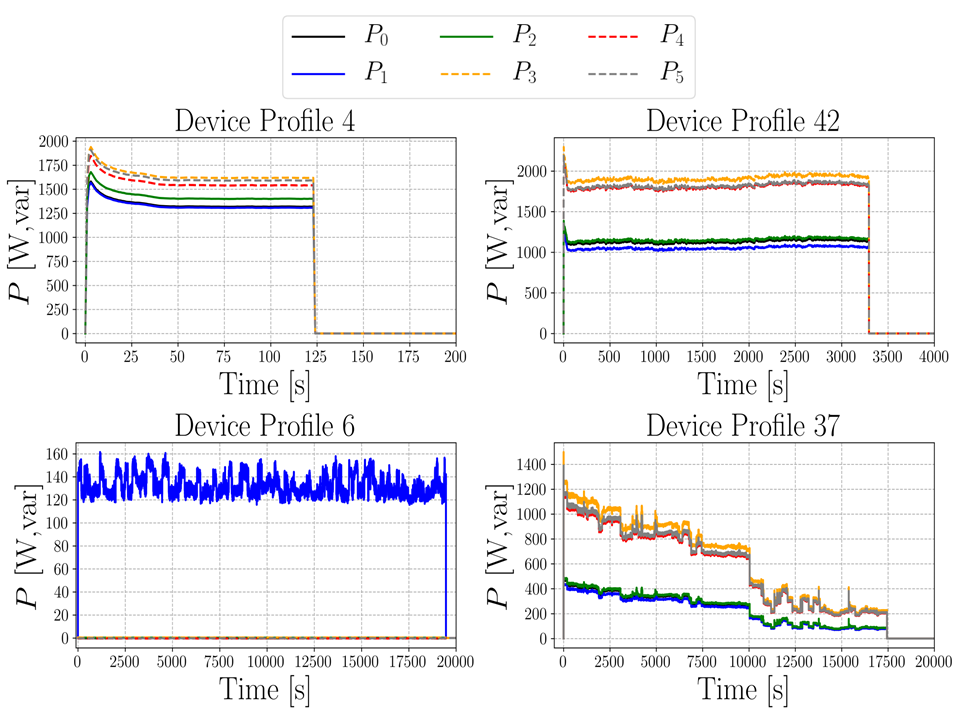}
\caption{Four examples of device profiles extracted unsupervised from the aggregate power signal. Solid lines represent active power and dashed lines represent reactive power. }
\label{fig:Signatures}
\end{figure}

Given the examples for different steps of the developed algorithm, we show four exemplary device profiles in Figure~\ref{fig:Signatures}. 
In total, we extracted 52 device profile from the aggregate power data with the developed algorithm.  
The depicted profiles are representative for all extracted device profiles since they show the main patterns and behaviors of the device profiles we extracted. 
Both upper illustrations in Figure~\ref{fig:Signatures} show the most frequent type of device profiles: A three phase connected device with a transient behavior in the beginning and afterwards a stable operating state where the relation between active and reactive power remains approximately the same. 
Additionally, Device Profile 4 and 42 show, that the relation of active and reactive power is characteristic for the specific device. 
The length of Profile 4 and 42 differs as well. 

Device Profile 6 shows no dynamic behavior in the beginning of the profile, but it consists of a constant component and an oscillating or random component. 
Obviously, Device Profile 6 represents a single-phase connected device, since all power values other than $P_1$ are close to zero. 
Profile 6 has a high ON-duration compared to Device Profile 4 and 42 with $d_6 \approx 20000~s$. 
An exception of the device profiles is represented by Device Profile 37, which shows a decreasing behavior with many little but sharp increases and decreases in all power features. 
This kind of device profiles is least frequent.

\subsection{Disaggregation}
\begin{figure}[!t]
\centering
\includegraphics[width=0.45\textwidth]{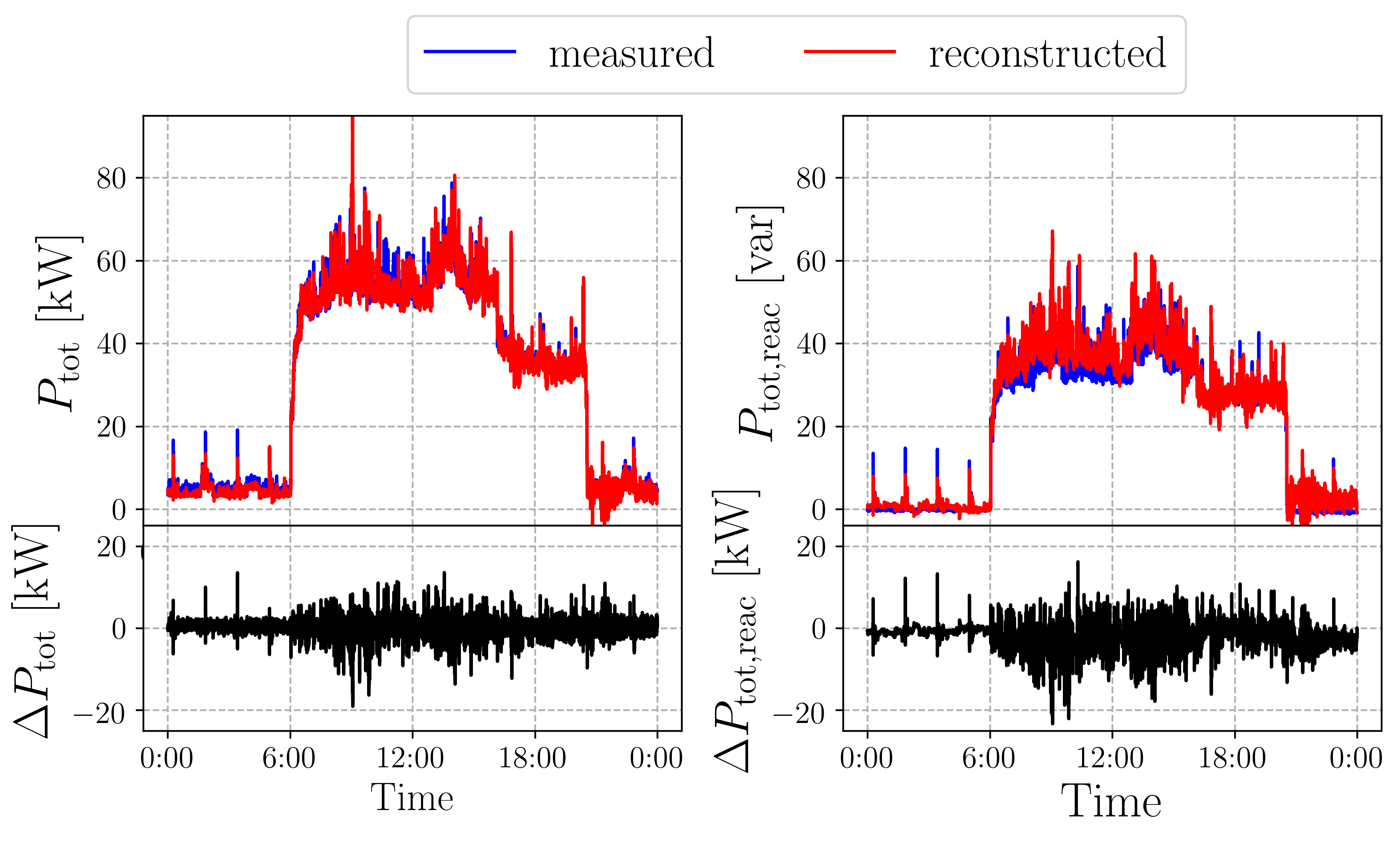}
\caption{Disaggregation results for 4\textsuperscript{th} December, 2018. On the left is the sum of active power shown and on the right side is the sum of reactive power shown. At the bottom, the respective absolute difference between measured and reconstructed power illustrated.}
\label{fig:Disaggregation_example}
\end{figure}

In total, we disaggregated power data of 119 days from December 2018 to March 2019. Figure~\ref{fig:Disaggregation_example} shows an exemplary day of data with the sum of active phases on the left and the sum of reactive phases on the right. 
Beneath, the respective absolute error is shown. 
It is visible, that the PSO is able to reconstruct the shape of the aggregate power signal over the duration of one whole day including the repetitive patterns during the night and most of the peaks. 
Nevertheless, there are error peaks of up to almost 20~kW which corresponds to approximately 25~\% of the measured power at the respective time.
Nevertheless, these high error values occur not frequently and they are of very short duration. 
During working time, the absolute error is higher than at night but there is no constant offset between measured and reconstructed power. 
The error of the reactive power is larger than the error of the active power. 
At the end of the presented day, noise is present in the reconstructed power. 

Table~\ref{tab:PSO_errors} shows the error values regarding all considered error measures for the working days of March 2019. 
The mean values of the daily evaluation of the reconstruction after disaggregation are presented as well as the respective standard deviation of the mean. 
The reproducibility of results is shown by the standard deviations of the mean error values which are approximately 10\% of the respective mean value. 
Especially, the daily consumed energy gets reconstructed very well with an average percentage error of less than one percent.  

\begin{table}[!t]
\renewcommand{\arraystretch}{1.2}
\centering
\caption{Error characteristics between the sum of active power measured and reconstructed after disaggregation for March 2019. The values are means of daily error evaluations and standard deviations of the mean values.}
\begin{tabular}{|c|c|}
\hline
  & 1 Month \\
& (March 1\textsuperscript{st}, 2019\\
& - March 31\textsuperscript{th}, 2019)\\
\hline
RMSE [W] 	& 1565 $\pm$ 150	\\
MAE [W] 		& 921 $\pm$ 85 	\\
MAPE [\%]			& 6.04 $\pm$ 1.51		\\
$\operatorname{Energy}_{\mathrm{E}}$[\%]		& 0.897 $\pm$ 0.156		\\
RMSE / Mean Power\,[\%] & 7.02 $\pm$ 0.67\\  

\hline
\end{tabular}
\label{tab:PSO_errors}
\end{table}

\subsection{Short Term Power Prediction}
Given the data produced by the disaggregation, an ANN is trained according to the in Section~\ref{chap:Methodology} described pre- and postprocessing of the data. The data set for testing the ANN performance consists of the 28\textsuperscript{th} and 29\textsuperscript{th} March, 2019. 
Therefore, we can calculate 188 power predictions of 15\,minutes each for the test data set. 
The ANN is given the first hour of the test set as input data. 
To put the results into perspective, we compare the error measures on the test set with error measures for different persistence forecasts. 
Lastly, we compare the developed short term prediction based on state changes data with prediction results of an ANN mainly based on past power data with a granularity of 5\,min from \cite{steens2020}. 
In \cite{steens2020}, the authors used the same data as in this work and optimized a Long-Short-Term-Memory Neural Network for a 24-h-day-ahead prediction. 
Although the prediction horizon and the granularity are different, the power prediction from \cite{steens2020} represents the standard prediction procedure and therefore acts as a benchmark prediction. 
All error measures are calculated for the sum of active power phases. 
Table~\ref{tab:persistence_error} shows the means and standard deviation of multiple error measures for the predictions. 
The first persistence forecast uses the power values from seven days ago whereas the second persistence forecast uses the power values of the preceding 15\,min. 
That means, for a prediction from $t_0 \dots t_0+900\,\mathrm{s}$ the power values from $t_0-900\,\mathrm{s} \dots t_0$ are taken. 
In comparison, Table~\ref{tab:ANN_error} shows the respective means and standard deviations of the error measures for the ANN predictions based on disaggregation data. 
The ANN outperforms both persistence forecasts regarding the mean error values of all calculated error measures. 
Especially, the MAPE and the error in daily consumed energy is significantly smaller. 

\begin{table}[H]
\renewcommand{\arraystretch}{1.2}
\centering
\caption{Multiple error measures between measured and predicted power for two different persistence forecasts. Presented are the means and standard deviations of the errors. They are calculated for 188 individual predictions of 15 minutes each for the test data set 28\textsuperscript{th} - 29\textsuperscript{th} March, 2019.}
\begin{tabular}{|l|l|l|}
\hline
 & Persistence & Persistence   \\
 & 7 days before & 15 min  \\
\hline
RMSE [W] 	& 6148 $\pm$	5755 & 4327 $\pm$	4045\\
MAE [W] 	& 5370 $\pm$	5762 & 3379 $\pm$	3722\\
MAPE [\%]	& 78.06	$\pm$ 117.84	&20.17 	$\pm$ 21.45	\\
$\operatorname{Energy}_{\mathrm{E}}$[\%]	& 35.25	$\pm$ 119.10&	3.40$\pm$	24.19\\
\hline
\end{tabular}
\label{tab:persistence_error}
\end{table}

\begin{table}[H]
\renewcommand{\arraystretch}{1.2}
\centering
\caption{Multiple error measures between measured and predicted power of the described ANN. Presented are the means and standard deviations of the errors. They are calculated for 188 individual predictions of 15 minutes each for the test data set 28\textsuperscript{th} - 29\textsuperscript{th} March, 2019.}
\begin{tabular}{|l|l|}
\hline
 & Power prediction\\
 & with ANN   \\
\hline
RMSE [W] 	& 3478 $\pm$ 3444	\\
MAE [W] 	& 2693 $\pm$ 3271 \\
MAPE [\%]	& 16.56 $\pm$ 20.19	\\
$\operatorname{Energy}_{\mathrm{E}}$[\%]	& -0.15 $\pm$ 22.47	\\
\hline
\end{tabular}
\label{tab:ANN_error}
\end{table}

\begin{figure}[!t]
\centering
\includegraphics[width=0.45\textwidth]{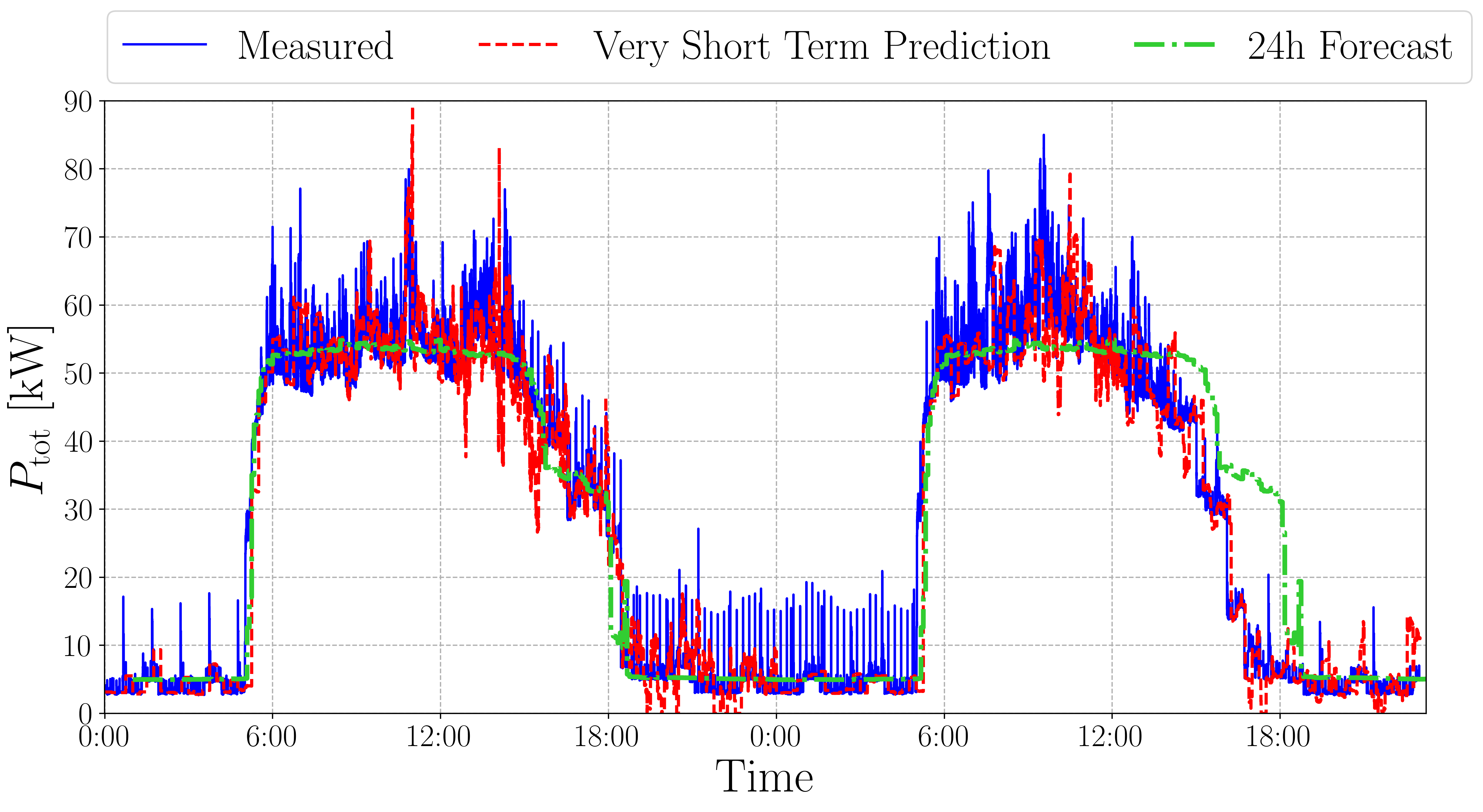}
\caption{Comparison of 24h-day-head power forecast and very short term power prediction based on state forecasts}
\label{fig:Forecast}
\end{figure}

The model from \cite{steens2020} is used to predict the power for the test set of this work. In Figure~\ref{fig:Forecast} the measured power and both ANN predictions are shown. 
The 24~h day-ahead prediction is similar to a rolling averaged power value whereas the short term prediction based on state changes data shows more the erratic behavior during working time with sharp increases and decreases in the power. 
For the model of the 24~h day-ahead prediction we can calculate the RMSE and MAE which results in RMSE $=$ 5124\,W and MAE $=$ 4507\,W. Both values are significantly higher than the mean error values of the short term prediction of the disaggregation based ANN.  

\section{Discussion}
\label{chap:Discussion}
For the extraction of device profiles, the main distinguishing factor is the behavior at an ON-event. 
The used peak criterion to determine the ON-behavior is very simple and neglects peaks with a width of multiple time-steps. 
This problem could be solved with a more sophisticated peak criterion in future work. 

The k-means clustering algorithm is used to determine clusters in the six-dimensional space of reactive and active power phases.
Also other publications use a clustering to differentiate between device types like \cite{Ng2009,Zeifman2012} but they use at maximum two features and not six features like in this work. 
In general, a clustering is more precise, the more characteristic features are present \cite{Bishop2006a}.
Thus, we can assume that we reach a higher precision in dividing the events into clusters of device types. 
Other properties measured by power analyzers could be used additionally to distinguish between different device types in future work. 
Nevertheless, the number of necessary features should be limited regarding a realistic application in real world energy management systems and the availability of high resolution power analyzers. 

During clustering, we assume that the cluster centers of OFF-events are the reversed cluster centers of ON-events.
When clustering is performed for ON-events and OFF-events individually, the OFF-event cluster centers with reversed signs lie within a 0.25-$\sigma$ area of the ON-event cluster center with $\sigma$ denoting the standard deviation of the respective cluster.
Therefore, the assumption can be justified.  
Additionally, Figure~\ref{fig:Centers} shows the symmetry to the central point of ON- and OFF-events. 
 
In order to determine the ON-event behavior we perform a peak analysis based on the assumption, that the switching procedure of a device is finished within one second. 
In reality, most devices show a transient behavior mostly in the shape of an exponentially decreasing oscillation \cite{Balzer2016}. 
Some publications distinguish between different transient behaviors and thus different device types \cite{leeb1995,chang2010}. 
But, compared to the measuring frequency, these processes happen on shorter timescales (within milliseconds) and can be neglected here. 
Only with a measuring frequency of kHz, the characteristic transient behavior would be observable \cite{Balzer2016}. 
But an exhaustive installation of measuring infrastructure which is able to measure in kHz is unlikely. 
Thus, the presented approach using a measuring frequency of 1~Hz is more realistic to be applied in local energy and power management systems. 
With the measuring frequency of 1\,Hz in this work an ON-event looks approximately like a step function in the aggregate power signal. 
Nevertheless, in most device profiles in Figure~\ref{fig:Signatures}, a transient behavior can be observed in the first few seconds of the according profiles. 
Thereafter, most devices reach a stable state for as long as the profile persists. 
Thus, the division of the device profiles in a stable state and a dynamic behavior for the stated formulation of the disaggregation problem can be applied here. 

The final step of the device profile extraction procedure is the median blending. 
In general, more samples to perform median blending with result in more precise results. 
Thus, it is very important to perform the extraction procedure on a sufficient amount of data. 
Especially, devices which are only switched on rarely result in less accurate profiles. 
The chosen normalization is carried out by means of a division by the maximum power value in every sample of $P_\mathrm{norm}$. 
With a high base load, this procedure could average out characteristic fluctuations of the device profile. 
Therefore, another normalization method might be appropriate if the individual profiles are of great importance and an allocation to real measured profiles is of interest.
However, we focused on the high quality very short term power prediction with the focus of attention on the aggregate power signal. 
Thus, small fluctuations of individual device profiles were of minor importance. 
The improvement of the median blending procedure or the application of other noise reduction methods for device profile extraction could be examined in future work. 

In total, 52 device profiles are extracted for our dataset of a commercial consumer. 
Since no additional knowledge of the used data is present, we can not validate this number of device profiles. 
But, the reason for this number of profiles is the division by ON-behavior and additionally the division of clusters into groups with similar run-time. Even a simple ohmic consumer type could therefore result in multiple device profiles.

A direct validation of the device profiles was not possible in this work due to a lack of data of the correct device profiles. Additionally, an assignment of extracted device profiles to measures, complex appliance signatures would be difficult since the extracted profiles only represent operational modes of appliances. 
But the good results in disaggregation and forecast show that the extracted device profiles are a satisfactory representation of the real device profiles.  

The extraction procedure has similarities to non-negative blind sources separation in acoustics, where the individual components and the the mixing procedure are unknown \cite{Pal2013}.
Since all methods used for the device profile extraction are from statistics and unsupervised machine learning, no hyperparameters have to be optimized to apply the algorithm to a different data set. 
The needed hyperparameters as the number of clusters $K$ or the number of Gaussian distributions in the GMMs are determined using statistical scores or criteria. 
Therefore, the device profile extraction algorithm can be applied without changes to other data sets. 
The transferability has to be examined systematically in the future. 

Figure~\ref{fig:Disaggregation_example} and Table~\ref{tab:PSO_errors} show, that the disaggregation of this work reaches a very accurate reconstruction of the measured power. 
The results are consistently good in all six phases. 
Thus, we can assume that the device profiles are a good representation of the real devices and also the separation according to the ON-event behavior seems valid. 
Since the PSO is a metaheuristic, incorrect assignments of devices to events are possible. 
Nevertheless, the disaggregation procedure produces additional knowledge of the building or the respective data set without a costly model building and adaption to the data. 
The aim of this work is the use of this additional knowledge for the purpose of a very short term power prediction and the examination if this additional knowledge provides benefits for such an application. 
The disaggregation procedure can be justified, if a disaggregation based prediction method outperforms classic prediction methods working in the power domain.  

The conducted very short term forecast using state changes data shows significantly better results than multiple persistence forecasts and a forecast using a LSTM network which is optimized for 24 hour prediction with a resolution of 5 min. 
Thus, the LSTM predicts 288 values compared to the 900 of our short term forecast. 
To be mentioned is, that the maximum accuracy of predictions is the accuracy of the reconstruction of the disaggregation. 
Thus, error values smaller than the reconstruction error values can only be undercut by chance, but not systematically.
The used model is a very simple ANN for a high number of input and target features. Thus, further optimization regarding the model of the neural network and maybe the use of LSTM layers or convolutional layers could result in better forecasts.
The ANN is optimized for the used data. 
Therefore, results could be worse, when applied to another data set of state changes. 
The developed forecast does not rely on a exhaustive rollout of measuring frequency as \cite{Alonso2020} and thus is easily transferable also with limited measuring infrastructure.  
Nevertheless, the transferability has to be examined systematically in the future.
 
It is to be assumed that a certain proportion of state changes of devices during working time is purely coincidental. 
However, randomness cannot be predicted by any model. 
In order to assess the chances of success of applying the presented approach to other power data, the randomness of the data could be determined in advance using appropriate methods. 
For example, the approximated entropy method described in \cite{pincus1991} could be used, which has already been applied to i.e. stock prices in~\cite{delgado2019}. 
Additionally, instead of a deterministic prediction, one could perform a probabilistic prediction and/or work with confidence intervals for the predicted power values. 
This procedure could help in management decision making.
In this work, we showed the advantages of the state changes data for power predictions. 
But the additional knowledge from device profile extraction and disaggregation could also be applied for other tasks like behavioral analysis, state analysis of the building, checking the health status of residents or employees or give recommendations for an intelligent power consumption regarding the availability of renewable energy. 
With more variable market-based electricity tariffs even new business models would be possible using the presented approach in energy management systems.

\section{Conclusions}
\label{chap:Conclusions}
In this work, we developed an algorithm for extracting device profiles from aggregate power data in six dimensions fully unsupervised. 
Since the method relies on statistical and unsupervised machine learning methods, it extracts repetitive patterns in the aggregate power data. 
Therefore, the extracted profiles are not necessarily full appliance signatures but one operational mode of one device. 
The direct validation of device profiles was not possible due to a lack of measured or correct device profiles. 
The transferability of the proposed device profile extraction is really high in theory, since no hyperparameters have to be optimized beforehand but this has to be proven in future work. 
The disaggregation uses the extracted device profiles and shows a very accurate reconstruction.
Thus, the device profiles seem to represent real appliance signatures sufficiently well. 
As the final application of the conducted NILM approach, the very short term prediction of power outperformed all compared predictions. 
Although, many publications developed or carried out various NILM algorithms, a broad application of those methods for other purposes is still missing. 
In this work, we showed the advantages of the additional knowledge of NILM for very short term power predictions. 
Our results and approaches for predictions could be combined with short term or long term power predictions working  directly in the power domain. 
Especially for energy management systems, such combined and high quality predictions would be very valuable for decision making processes.

\appendix
\section*{Acknowledgments}
The authors acknowledge the financial support of the Federal Ministry for Economic Affairs and Energy of the Federal Republic of Germany for the project \textit{EG2050: EMGIMO: Neue Energieversorgungskonzepte für Mehr-Mieter-Gewerbeimmobilien (03EGB0004G and 03EGB0004A)}. For more details, visit www.emgimo.eu. The presented study was conducted as part of this project.

\end{document}